\def\bq{\begin{equation}}
\def\eq{\end{equation}}
\def\bqa{\begin{eqnarray}}
\def\eqa{\end{eqnarray}}
\def\bqb{\begin{eqnarray*}}
\def\eqb{\end{eqnarray*}}
\documentstyle[12pt]{article}\textwidth 16cm
\def\pr#1#2#3{ Phys. Rev. ${\bf{#1}}$ (#2) #3 }
\def\prl#1#2#3{ Phys. Rev. Lett. ${\bf{#1}}$ (#2) #3 }
\def\pl#1#2#3{ Phys. Lett. ${\bf{#1}}$ (#2) #3 }
\def\prep#1#2#3{ Phys. Reports ${\bf{#1}}$ (#2) #3 }
\def\np#1#2#3{ Nucl. Phys. ${\bf{#1}}$ (#2) #3 }
\def\zp#1#2#3{ Z. Phys. ${\bf{#1}}$ (#2) #3 }

\def\ie{{\it i.e.\/}}
\def\eg{{\it e.g.\/}}

\def\etal{{\it et.al.\/}}

\def\L{ {\cal L }}
\def\O{ {\cal O }}

\global\nulldelimiterspace = 0pt

\def\roughly#1{\mathrel{\raise.3ex
    \hbox{$#1$\kern-.75em\lower1ex\hbox{$\sim$}}}}
\def\lsim{\roughly<}
\def\gsim{\roughly>}

\def\op#1{{\widehat #1}}

\def\ol#1{\overline{#1}}

\def\mwd{M_W^2}
\def\mw{M_W}
\def\lw{\lambda_W}

\def\Uh{\widehat{U}}

\begin{document}
\pagenumbering{arabic}
\thispagestyle{empty}
\def\thefootnote{\fnsymbol{footnote}}
\setcounter{footnote}{1}

\begin{flushright} PM/94-29 \\ THES-TP 94/07 \\
July 1994 \\
corrected version \end{flushright}
\vspace{2cm}
%---------------------titre ---------------------------------------
\begin{center}
{\Large\bf Dynamical Scenarios for $SU(2)_c$ Symmetric New
Physics Interactions}\footnote{Work supported
by the scientific cooperation program between
 CNRS and EIE.}
 \vspace{1.5cm}  \\
%-----------------------------------------------------------------
{\large G.J. Gounaris$^a$, F.M. Renard$^b$ and G.
Tsirigoti$^a$}
\vspace {0.5cm}  \\
$^b$ Physique
Math\'{e}matique et Th\'{e}orique,
CNRS-URA 768\\
Universit\'{e} Montpellier II,
 F-34095 Montpellier Cedex 5.\\
\vspace{0.2cm}
$^a$Department of Theoretical Physics, University of Thessaloniki,\\
Gr-54006, Thessaloniki, Greece.\\

\vspace {2cm}

 {\bf Abstract}
\end{center}
\noindent
We study a picture of effective
interactions among the $W$ and Higgs bosons which
is consistent with the precision
tests at present energies, and at the same time allows for
large observable New Physics (NP) effects in the bosonic sector.
Toy dynamical models containing new heavy particles
are used to indicate how such a picture could be created by
integrating out the heavy particles. In these models custodial
$SU(2)_c$ symmetry is realized at a certain large scale
$\Lambda_{NP}$, either naturally or at a certain limit of a
single coupling. In the derivation
of the effective lagrangian we keep operators of dimension up to
six. These operators involve only the known gauge boson and
Higgs particles. They provide a valid description of NP up to an energy
scale which is just below the mass of the new heavy degrees
of freedom.\\

\vspace{3cm}

\def\thefootnote{\arabic{footnote}}
\setcounter{footnote}{0}
\clearpage

Signals of New Physics (NP) beyond the Standard Model (SM) could
appear in two different ways. The first possibility is that
new particles associated with NP will be discovered at the
future colliders. This spectacular case should appear
at a future collider whose energy is larger than the
typical mass of the new particles associated with NP. Standard
arguments \cite{Chan} suggest that this
energy is in the TeV range,
although it is not excluded that low lying
states (like \eg\@ the lightest Higgs, the lightest
supersymmetric particle, or new
vector bosons) could be found earlier.\par

 The second possibility
is that all observable new particles are so heavy, that they
cannot be directly produced by the future colliders.
In such a case the
only possibility left for an NP discovery is through
precision tests probing departures from the SM
predictions for processes involving usual
particles. Such tests have already been performed for fermionic
interactions at LEP1, SLC and in low energy experiments
\cite{Mars}.
They indicate that the effective lagrangian describing NP should
not include
(at least at the few permille level) \underline{light}
fermion fields.
The \underline{heavy} quark sector
has not yet been tested with such an accuracy though.
For example, the $Z\to b \bar b$
width and the $b \bar b$ asymmetries just approach the percent
level \cite{CRV}, while the consistency of top quark properties
with the $SM$ expectations are still to be confirmed at this
level of accuracy \cite{top}.
On the other hand the bosonic sector (gauge boson self-interactions
as well as the whole scalar sector) have not yet been directly
tested at all.\par

It is generally expected (and hoped) that NP will partially
reveal the mechanism responsible for generating the masses
of the SM particles. Thus, it should not be surprising if
NP signals first appear in  sectors  where mass generation
plays an important role, like \eg\@ the heavy fermionic
or the bosonic sector. From the
agreement between the measured $Z$ couplings to light fermions and
the SM predictions, one concludes that if higher vector
bosons exist and couple to light fermions with a standard
electroweak strength, then their masses should lie at or above the 1
TeV range \cite{Zprime}. This limit lowers to a few hundred of
GeV, if the new
vector bosons couple very weakly to the light
leptons and quarks \cite{Bilenky}. Concerning the Higgs particle,
no serious indication on its
mass range (or to some extent even its existence) is yet
provided, apart from a lower mass limit
of about 60 GeV \cite{Higgs}. \par

This leaves room for many possibilities for the appearance of
NP, which  include
the cases demanding that the Higgs boson is replaced by a
strongly interacting sector at the TeV scale \cite{Chan, BESS}.
Thus in the present paper, we concentrate in a
framework where
all new degrees of freedom are very heavy, and the interactions
among the known particles up to a scale
$\Lambda_{NP} \gsim 1 $ TeV, are described by an effective
lagrangian satisfying
$SU(2)\times U(1)$ gauge symmetry and consisting of
the SM lagrangian, plus additional NP operators involving only
the known gauge boson and Higgs fields. The order of magnitude of the
contributions of these NP operators is
determined by the scale $\Lambda_{NP}$. To lowest
nontrivial order we restrict to NP operators of dimension up to
six \cite{Buchmuller}.   \par

These NP operators induce both direct and indirect contributions
to the 2-point gauge boson functions ($\gamma$, $Z$, $W$
self-energies).
The direct ones appear at the tree level
and they are strongly suppressed due to the present precision
measurements. The indirect contributions though, induced
by the use of the effective
interactions inside loops,
must be handled with care in order to be meaningful.  Reliable
constraints can then only be
obtained from effective lagrangians which lead to a decent high
energy behaviour and do not produce results strongly depending
on the scale $\Lambda_{NP}$; \ie\@ on the details of the regime
above or close to the NP threshold. Such a behaviour can best be achieved
if the NP operators satisfy $SU(2)\times U(1)$ gauge invariance
\cite{DeR}, which is of course
mandatory for any decent theory. In this respect it is
interesting to mention that
it is always possible to write any
interaction respecting electromagnetic gauge invariance,
in a $SU(2)\times U(1)$ gauge invariant form,
provided no constraint on the dimensionality of the allowed operators
is imposed \cite{London, GR1}.  So $SU(2)\times U(1)$ gauge invariance
alone does not restrict the kind of possible interactions,
since the implied improvement of the high energy
behaviour can always be achieved either by the
additional multi-boson vertices
or by vertices involving Higgses.
 However, if $\Lambda_{NP} \gg v$ ($v$ is the
electroweak scale), restrictions will be provided by the
dimension of the effective operators.\par

 In addition to  $SU(2) \times U(1)$
gauge invariance,  in the present paper we
assume that NP satisfies CP conservation and restrict to
purely bosonic operators of dimension up to six. This assumption
is motivated by remarking that the fermionic contribution to
the $SU(2)\times U(1)$ currents may always be eliminated by
using the gauge boson equations of motion \cite{Buchmuller},
while the remaining fermionic terms should somehow be related to
the mechanism responsible for the spontaneous symmetry breaking.
Because of this, their strength may be similar to the Yukawa
type fermionic couplings which are known to be small, except
for the top quark case. Thus, following \cite{Hagi} we disregard
all fermionic operators and describe NP in terms
of boson operators only. The equations of motion are then used
only so far they relate bosonic operators among themselves.
The a priori very long list of these
operators reduces to only 11 independent ones.\par

Four of these operators may be written as,
\bqa
\overline{\O}_{DW} & =&  4 ~ \langle ([D_{\mu} , W^{\mu \rho}])([D^{\nu} ,
W_{\nu \rho}]) \rangle \ \ \
\ \ \ \ \ \ \ \ \ \  , \ \ \ \\[0.3cm]
\O_{DB} & = & (\partial_{\mu}B_{\nu \rho})(\partial^\mu B^{\nu
\rho}) \ \ \ \ \ \ \ \ \ \ \ \ \ \ \ \ \ \ \ \ \ \ \
\ , \ \ \ \\[0.3cm]
\O_{BW} & =& \Phi^\dagger B_{\mu \nu} W^{\mu \nu} \Phi
\ \ \ \ \ \ \ \ \ \ \ \ \ \ \ \ \ \ \ \ \ \
 \ \ \ \ \ \ \ , \ \ \\[0.3cm]
\O_{\Phi 1} & =& (D_\mu \Phi)^\dagger \Phi \Phi^\dagger
(D^\mu \Phi) \ \ \ \ \ \ \ \ \ \ \ \ \ \ \ \ \ \ \ \ \ . \ \ \
\eqa
In addition to these we also give for later convenience the
redundant operator\footnote{The
operator $\O_{\Phi 4}$ of (5) was first introduced in \cite{Buchmuller}.
It was then omitted in \cite{Hagi} because partial
integration of $\O_{\Phi 2}$
(see (9)), leads to
$\O_{\Phi 2} +8 \O_{\Phi 4}+4 (\Phi^\dagger \Phi)\left\{
(D_\mu D^\mu \Phi^\dagger)\cdot \Phi
+\Phi^\dagger \cdot (D_\mu D^\mu \Phi) \right \} =0$ , which
combined with the Higgs equations of motion (where the Yukawa
contributions are neglected) implies that $\O_{\Phi
4}$ is equivalent to a combination of purely
Higgs operators ~without
any gauge boson couplings. Note that the substitution of
$\O_{\Phi 4}$ by these Higgs fields implies a renormalization of
the parameters already existing in SM, which induces the
necessary changes to the gauge couplings ~explicitly modified by
the presence of $\O_{\Phi 4}$.}
\bqa
\O_{\Phi 4} & =& (\Phi^\dagger \Phi)(D_\mu \Phi)^\dagger(D^\mu
\Phi) \ \ \ \ \ \ \ \ \ \ \ \ \ \ \ \ \ \ \ \ \ . \ \
\eqa
All these operators contribute directly to the gauge boson
2-point functions, and
according to the precision tests, they should have reduced couplings.
In the preceding equations, the definitions \\
\bq \Phi=\left( \begin{array}{c}
      \phi^+ \\
{1\over\sqrt2}(v+H+i\phi^0) \end{array} \right) \ \ \ \ , \ \
\eq
\bqa
D_{\mu} & = & (\partial_\mu + i~ g_1 Y B_\mu +
i~ g_2 W_\mu ) \ \ \ \ , \ \ \nonumber
\eqa
\bq
W_\mu = \overrightarrow{W}_\mu \cdot
\frac{\overrightarrow{\tau}}{2} \ \ \ \ \ , \ \ \ \ \ \ \
W_{\mu \nu}= \overrightarrow{W}_{\mu \nu} \cdot
\frac{\overrightarrow{\tau}}{2} \ \ \ \ \ , \ \
\eq
are used, where $Y$ gives the hypercharge of the field to which
the covariant derivative acts, $v$ is vacuum expectation value
of the Higgs field and $\langle A \rangle \equiv TrA$.
For studying the  $SU(2)_c$ transformations
of the NP operators, we quote the
expression for the Higgs field\\
\bq
\Uh=\bigm(\widetilde \Phi\ \ , \ \Phi\bigm) \ \ \ \
, \eq
where $\widetilde \Phi = i\tau_2 \Phi^* $.\par

Another two NP operators, constructed with scalar fields only,
are
\bqa
\O_{\Phi 2} & = & ( \partial_\mu \langle \Uh \Uh^\dagger \rangle )
(\partial^\mu \langle \Uh \Uh^\dagger \rangle ) \ \ \ \ \ \ \ , \ \ \
\\[0.3cm]
\O_{\Phi 3} & = &  \langle \Uh \Uh^\dagger \rangle ^3\ \ \ \ \ \ . \ \
\eqa
We call them "superblind", since they do not give observable
contributions to the present precision measurements even at the
one-loop level.\par

Finally the last five operators
\bqa
\O_W &= & {1\over3!}\left( \overrightarrow{W}^{\ \ \nu}_\mu\times
  \overrightarrow{W}^{\ \ \lambda}_\nu \right) \cdot
  \overrightarrow{W}^{\ \ \mu}_\lambda =-{2i\over3}
\langle W^{\nu\lambda}W_{\lambda\mu}W^\mu_{\ \ \nu}\rangle \ \ \
, \ \ \ \ \\[0.3cm]
\widehat{\O}_{UW} & = & \frac{1}{2}\, \langle \Uh\Uh^{\dagger}
\rangle  \langle W^{\mu\nu} \
W_{\mu\nu}\rangle \ \ \ \ \ \ \ \ , \ \ \ \\[0.3cm]
\widehat{\O}_{UB} & = & \langle \Uh\Uh^{\dagger} \rangle B^{\mu\nu} \
B_{\mu\nu} \ \ \ \ \ \ \ \ \ , \ \ \ \\[0.3cm]
\O_{W\Phi} & = & 2~ (D_\mu \Phi)^\dagger W^{\mu \nu} (D_\nu \Phi) \ \ \ \ \
\ \ \ \ \ , \ \  \ \\[0.3cm]
\O_{B\Phi} & = & (D_\mu \Phi)^\dagger B^{\mu \nu} (D_\nu \Phi)\ \ \ \ \
\ \ \ \ \ \ , \ \  \
\eqa
constitute the most interesting set since the present precision
experiments still allow them to be appreciable.\par

Note that in this list of independent operators,
we have used  $\overline O_{DW}$ given in (1),
instead of the one usually called $O_{DW}$ defined as
\bq
\O_{DW}  =  2 ~ \langle ([D_{\mu} , W_{\nu \rho}])([D^{\mu} ,
W^{\nu \rho}]) \rangle \ \ \ \ \ , \ \ \ \
\eq
and  related to $O_W$ and $\overline{\O}_{DW}$ by the
identity
\bq
 \O_{DW} ~= ~ 12~g_2 \O_W + ~ \overline{\O}_{DW}\ \ \  \ \ \
\ \ \ \ .\ \ \ \ \ \ \ \
\eq
The reason is the following. From the expressions of
these operators given in (11, 1, 16) we
observe that $\overline O_{DW}$ and $\O_{DW}$
generate the same contribution to the 2-point gauge function
at the tree level. The difference between these operators first
arises in the triple gauge vertices. A study
reveals that the triple gauge vertex in $\overline O_{DW}$
has the usual Yang-Mills form (apart from additional d'Alembertien
derivatives), whereas $\O_{DW}$ includes in addition the
genuinely anomalous 3-gauge quadruple coupling described by
$O_W$ \cite{Kuroda}. This later coupling does \underline{not}
appear in $\overline O_{DW}$.\par

The five operators appearing in (11-16) are called
"blind" \cite{DeR}, since they contribute  to the
quantities measured in the precision experiments at
LEP1, only at  the one-loop
level\footnote{For the peculiarities of $\widehat{\O}_{UW}$ and
$\widehat{\O}_{UB}$ see \cite{GLR} and below.}.
As one could imagine from a rough estimate based on the
$(\alpha/4\pi)$ loop factors, these measurements only produce mild
indirect constraints on the couplings of these operators.
 This is what comes out when effective
operators are treated one by one.
 If several operators are
simultaneously considered though, cancellation effects appear and
even these mild constraints essentially vanish \cite{Hagi}.
Because of this, present measurements allow the
couplings of the blind operators to be considerable, which in
turn makes the operators themselves interesting.\par

The energy domain where such operators can appear as local,
has also been discussed on the basis of unitarity
considerations \cite{GLR, GLPR}. Locality,
together with high dimensionality,
will inevitably produce amplitudes which  approach
the unitarity limit at a sufficiently high scale. This phenomenon
is well known from the old Fermi current-current interaction,
which is also a $dim=6$ local operator. As in the Fermi theory, these
unitarity constraints give indications not only on the domain of
validity of the effective theory, but also on the threshold of
the related NP. Thus,
depending on the type of operators, at sufficiently high
energies either strong
interactions will appear involving gauge bosons of various
helicities, or new degrees of freedom will be excited\footnote{In fact
this later possibility would be analogous to the Fermi case,
where the $W$ excitation destroyed
the locality of the current interaction.}.
The scale where this happens can be identified as
$\Lambda_{NP}$.
Relations between the NP coupling constants and the saturation scales
$\Lambda_{NP}$ have been established for each operator
\cite{GLR,GLPR}. Assuming that
$\Lambda_{NP}$ is of the order of 1 TeV, one then gets
"unitarity" upper bounds for
these "anomalous" couplings, which are as stringent as those
obtained indirectly from the LEP1
measurements. Alternatively, if an
upper bound on any of these couplings is experimentally
established, then the unitarity relations may be used to give a
lower bound on the threshold of the related NP.\par

To summarize the present situation concerning the five "blind"
operators, we state that LEP1 and unitarity
do not strongly constrain them. In principle much stronger constraints
will be obtained from direct measurements on these operators.
One has of course to check if this is achievable in the
contemplated
future experiments. In this respect we note that at LEP2
\cite{LEP2},
couplings of the above "blind" operators at the level of
at least $O(0.1)$ are expected to be visible.
Subsequently, LHC \cite{GR2, GLRLHC},  and NLC
\cite{NLC, GLRNLC} should allow to reach the $O(0.01)$
level, or even the permille level, for such NP couplings.
In addition, laser
back scattering experiments may turn out to be especially stringent
for studying \eg\@ the $H\gamma \gamma $ anomalous (and
normal) coupling, in case a light Higgs boson could be produced
through the
$\gamma\gamma$ collisions \cite{Baillargeon, GLRHiggs}. So
finally the same level of accuracy
should be reached for the bosonic sectors at NLC, as for the
fermionic sector at LEP1. It makes therefore sense to discuss
the NP effects in these various sectors and from their
comparison, to try to infer the NP properties.\par

The fact that no NP effect has been seen in the
fermionic sector at LEP1, and the possibility that such
an effect could be seen in the
bosonic sectors, should not be considered as an "unnatural"
situation, but rather as a
remarkable signal of the NP properties. Like \eg\,  the
presence of a certain symmetry.
 An example of such symmetry is
custodial $SU(2)_c$.
Cumulating the information that
lepton and quark couplings seem standard, and that
neither W-B mixing nor
$\Delta\rho$ effects in the self-energy
$\epsilon $  parameters  appear in nature, provides
ample motivation for the
possible importance of this global symmetry for NP \cite{Sikivie,
GR2, GLR}. This motivation is further supported  by the
expectation that NP may be related to the scalar
sector of the electroweak interactions, which is known to
preserve $SU(2)_c$ in SM, whereas the $B_{\mu}$ and fermionic
couplings violate it. \par

An $SU(2)_c$ preserving NP effective
lagrangian should include neither ordinary fermions nor the $B_{\mu}$
fields. It can only involve the $W_{\mu}$ and Higgs
fields\footnote{ This use of
$SU(2)_c$ symmetry is somewhat stronger than the one where it is
applied only after the $B_{\mu}$
field has been removed ($ g_1 \to 0$) \cite{Sikivie}.}.
If we assume that NP somehow respects global
$SU(2)_c$ symmetry, then
the above set of 11 operators is restricted to only 5 ones namely,
the two
superblinds $\O_{\Phi 2}$ , $\O_{\Phi 3}$,
the two operators $O_W$ and $\overline O_{DW}$ involving $W_\mu$
only, and
the operator $\widehat{O}_{UW}$ involving both $W_\mu$ and
Higgs fields.\par

At this point it is worthwhile to offer examples of dynamical
models for NP which at the effective lagrangian level
have to some extent the $SU(2)_c$ symmetry mentioned above.
Thus in model A below, the effective NP
interaction at the large scale $\Lambda_{NP}$ obeys global $SU(2)_c$
invariance in the limit that a certain coupling called
$g_{\psi2}$ vanishes, while in models B and B$^\prime$ this
symmetry is realized independently of the actual values of the
couplings. At a lower scale the couplings of these
operators will of course run according to
the rules implied by the SM lagrangian. Consequently smaller
contributions from $SU(2)_c$ violating operators will also be
generated at such lower scales. \par

\vspace{0.5cm}
\noindent
\underline{Model A:}\par
In this model, we assume that NP is determined by a complex scalar
field $\Psi$
which under the $SU(2)\times U(1)$ gauge group has
isospin $I$
and vanishing hypercharge. The associated $\Psi$
particles are assumed to have a large
mass $M=\Lambda_{NP}$, and possibly also some hyper-colour
$\tilde{N}_c$. The basic renormalizable lagrangian will then
be the sum

\bq \L=\L_{SM}+\L_{\psi} \ \ \ \ \ \ , \eq
of the SM lagrangian
\bqa
\L_{SM}&=& -{1\over 2}\langle W_{\mu\nu}W^{\mu\nu}\rangle-
{1\over4}
  B_{\mu\nu}B^{\mu\nu} +\frac{1}{2}\langle D_\mu
\Uh D^\mu \Uh^{\dagger} \rangle\nonumber\\[.3cm]
 \null & \null & -{ M^2_H\over 8v^2}
\left ( \langle \Uh \Uh^{\dagger}\rangle-v^2\right)^2+\
 \makebox{fermionic terms \ \ \ \ , }
\eqa
and the lagrangian describing the new degrees of freedom
\bq
\L_{\psi}= (D_\mu \Psi )^\dagger(D^\mu \Psi
)-\Psi^\dagger \left (\Lambda_{NP}^2 -g_{\psi 1}\langle
\Uh \Uh^\dagger \rangle
-g_{\psi 2} \Uh \tau_3
\Uh^\dagger \right ) \Psi \ \ \ \ \ ,  \ \ \
\eq
where the definition (8) has been used.
The tensorial representation of the isospin=I field $\Psi$ is
implicitly used,
so that \eg\ in the last term of (20) one of the  isospin
indices of $\Psi$ and
$\Psi^\dagger$ are dotted  to $\Uh\tau_3 \Uh^\dagger$, while
the remaining indices are dotted among themselves. This term
violates $SU(2)_c$. We note that
(20) gives, apart from an irrelevant $(\Psi^\dagger \Psi)^2$
term, the most general renormalizable interaction in consistence with
$SU(2)\times U(1)$ gauge invariance. The vanishing
hypercharge of $\Psi$ implies that no $B_\mu$ interactions
appear in (20).  \par

The standard techniques \cite{Georgi} may now be used to calculate
the effective lagrangian describing the electroweak interactions
at a scale $\mu$ just below $\Lambda_{NP}$. This is obtained by
integrating out at the one loop order the heavy degrees of
freedom described by the field $\Psi$. Thus, at the scale
just below $\Lambda_{NP}$,
the effective electroweak interaction
among the usual SM particles,
is obtained by employing the  Seeley-DeWitt expansion of the
relevant determinant\footnote{
This result has been also checked by an explicit calculation of
the self energy and triangle loops.}
\cite{Ball}. It is given by
\bq
\L_{eff}= \L_{SM} + \L_{NP} \ \ \ \ \ \ \ \ \ ,\ \ \ \ \ \ \
\eq
\bqa
\L_{NP} & = &\frac{(2I+1) \widetilde{N}_c}{(4\pi)^2}~
\Bigg\{-~g_{\psi 1} \Lambda^2_{NP} \left (\frac{1}{\epsilon}+1
\right )\langle \Uh
\Uh^\dagger \rangle +
\frac{1}{2 \epsilon}\left (g^2_{\psi 1} +g^2_{\psi 2}~
\frac{I(I+1)}{3}\right ) \langle \Uh \Uh^\dagger \rangle^2
 \nonumber\\[0.3cm]
\null & \null &
- ~ \frac{g^2_2 I(I+1)}{18 \epsilon}\langle W_{\mu
\nu} W^{\mu \nu} \rangle
+ ~ \frac{1}{6 \Lambda^2_{NP}}(g^3_{\psi 1} + g_{\psi 1}
g^2_{\psi 2}I(I+1))
\langle \Uh \Uh^\dagger \rangle ^3 \ \ \nonumber\\[0.3cm]
\null & \null &
+ ~ \frac{1}{12 \Lambda^2_{NP}}\left [
\left (g^2_{\psi 1} +g^2_{\psi 2}\frac{I(I+1)}{3}\right ) \O_{\Phi 2}
{}~+~ g^2_{\psi 2}\frac{16 I(I+1)}{3} (\O_{\Phi 4} -\O_{\Phi 1})
\right ]\ \ \nonumber\\[0.3cm]
\null & \null &
- ~ \frac{g^2_2 I(I+1)}{90\Lambda^2_{NP}}\left [
 10 g_{\psi 1} \widehat{\O}_{UW}
+  g_2 \O_W + \frac{1}{4}\, \overline{\O}_{DW}
\right] \Bigg\} \ \ , \
\eqa
where the definitions (1,4,5,8,9,11,12) have been used, and $\epsilon
=(2-n/2)$ is the usual dimensional regularization parameter for
the ultraviolet divergences.\par

The first three terms in (22) renormalize  $SU(2)_c$
invariant quantities already existing
in the SM lagrangian (19), while the remaining ones create
contributions from the five possible $dim=6$, $SU(2)_c$
conserving   operators  $\O_{\Phi 2}$ , $\O_{\Phi 3}$,
$O_W$, $\overline O_{DW}$ and $\widehat{O}_{UW}$.
Finally $\O_{\Phi 4}-\O_{\Phi 1}$ gives the only $SU(2)_c$
violating but $B_\mu$ independent
 contribution, which is generated by the $g_{\psi 2}$ coupling.
The fact that no $B_\mu$ couplings appear in (22) is a direct
consequence of the vanishing hypercharge of the $\Psi$ field.\par

The term $W_{\mu \nu}W^{\mu \nu}$ in (22) induces
a non observable renormalization of $W_\mu$, while
the purely Higgs operators
renormalize the vacuum expectation value
and the mass of the Higgs field, producing in addition
anomalous self-interactions. After this
renormalization is done we are lead to
\bq
\L_{NP}=\frac{2 \tilde{d}}{v^2}~ \widehat{\O}_{UW}+
\frac{4 \tilde\lw }{g_2 v^2}~\O_W +\frac{4 \tilde x_{\ol{DW}}}{g^2_2 v^2}~
\ol{\O}_{DW}+ \frac{g_\phi}{\Lambda^2_{NP}}(\O_{\Phi 4} -
\O_{\Phi 1}) +
\makebox{(Higgs self-interactions)}\ .
\eq
Using (21,22) and $v=2\mw / g_2$, we get
\bq
\tilde \lw =- ~\frac{(2I+1)I(I+1)\widetilde{N}_c}{90\ (4\pi)^2}~
\frac{g^2_2 \mwd}{\Lambda^2_{NP}} \ \ \ \ \ , \ \ \ \
 \eq
\bq
\frac{\tilde x_{\ol{DW}}}{\tilde \lw}~=~\frac{1}{4} \ \ \ \ \ \ , \ \ \ \
\eq\\
\bq
\frac{\tilde d}{\tilde \lw}~=~ \frac{20\  g_{\psi 1}}{g^2_2}\ \ \ \
. \ \ \ \
\eq
Eliminating now the trivial contribution to the $W$ kinetic
energy from $\widehat{\O}_{UW}$ by renormalizing the $W_\mu$
field and $g_2$ as in \cite{GLR} we get
\bq
\L_{NP}=d \O_{UW}+
\frac{\lw g_2}{\mwd}~\O_W +\frac{ x_{\ol{DW}}}{\mwd}~
\ol{\O}_{DW}+\frac{g_\phi}{\Lambda^2_{NP}}(\O_{\Phi 4} -
\O_{\Phi 1})+
\makebox{(Higgs self-interactions)} ,
\eq
where \cite{GLR}
\bq
\O_{UW}~=~\frac{1}{v^2} \, \langle \Uh \Uh^\dagger - \frac{v^2}{2}
\rangle \langle W_{\mu \nu}W^{\mu \nu} \rangle \ \ \ \ \ , \ \ \
\eq
and
\bq
d~=~ \frac{\tilde d}{1-2 \tilde d} \ \ \ \ \ , \ \ \ \
\eq
\bq
\lw~=~ \frac{\tilde \lw}{(1-2\tilde d)^2}\ \ \ \ ,\ \ \ x_{\ol{DW}}~=~
\frac{\tilde x_{\ol{DW}}}{(1-2\tilde d)^2} \ \ \ , \  \ \ \
\eq
satisfying again
\bq
\frac{ x_{\ol{DW}}}{\lw}~=~\frac{1}{4}\ \ \ \ \ \ \ . \ \ \ \ \
\eq\par

Note that the difference between the tilded couplings in (24-26)
and those without it in (28-30), arises from the respective use of
$\op{\O}_{UW}$ and $\O_{UW}$ in $\L_{NP}$; compare (23,
27). To lowest order
in the electroweak factor $g^2_2/(4\pi)^2$ this difference vanishes.
The  magnitude of these couplings is controlled by
$g^2_2/(4\pi)^2$, the $dim=6$ scale ratio $\mwd/\Lambda^2_{NP}$
and the isospin and hyper-colour coefficients. Note the ratio of
$4/1$ in favour of $\O_W$ over $\ol{\O}_{DW}$, and the negative
signs of both couplings. Since $\ol{\O}_{DW}$ is already
strongly constrained by LEP1 \cite{Hagi}, this result provides
only little chance for the observability of $\lw$.\par

One also sees that the three Higgs involving $SU(2)_c$
invariant operators
$\O_{\Phi 2}$, $\O_{\Phi 3}$ and $\O_{UW}$
owe their presence to the basic coupling $g_{\psi 1}$ in (20).
Most interesting is the operator
$\O_{UW}$ whose physical consequences have been studied in
\cite{GLRLHC, GLRNLC, GLRHiggs}.
Contrary to the case for the operators $\O_W$ and $\ol{\O}_{DW}$,
the coupling of $\O_{UW}$ could not be predicted even if
$\Lambda_{NP}$ were known, because $g_{\psi 1}$ is
unknown. Nevertheless it is interesting to remark that
according to (26,29,30),
$d$ may be considerably larger than $\lw$ if $g_{\psi
1}$ is comparable to the electroweak coupling $g^2_2$.
Finally the only $SU(2)_c$ violating
contribution in this model arises from $\O_{\Phi 4}-
\O_{\Phi 1}$ and is determined by the other
unknown coupling $g_{\psi 2}$.  The single direct contribution of
this operator is to $\Pi_{WW}$, which in turn implies that the only
precision measurement parameter which is sensitive to it
is $\epsilon_1 = \Delta \rho $. Present precision measurements
constrain $g_{\psi 2}$ to be very
small. \par

\vspace{0.5cm}
\noindent
\underline{Model B:}\par
We now try a second toy model which naturally generates only $SU(2)_c$
invariant effective interactions. In this model NP is given by a complex
fermion field $F$ whose left and right components have both
vanishing hypercharge, isospin $I$ and hyper-colour $\widetilde{N}_c$.
The associated particles again have a large  gauge invariant
mass $M_f=\Lambda_{NP}$. The basic lagrangian is now
\bq
\L= \L_{SM} + \L_F \ \ \ \ \ \ \ \ \ ,\ \ \ \ \ \ \
\eq
where (19) is used and the most general $SU(2)\times U(1)$ gauge
invariant and renormilazable lagrangian for NP is
\bq
\L_F=~i \ol{F} (\rlap/\partial +i g_2 \rlap /{\overrightarrow
W} \cdot \overrightarrow{t} )F
-\Lambda_{NP} \ol{F}F \ \ \ \ \ ,  \ \ \
\eq
with $\overrightarrow{t}$ denoting the isospin $I$ matrices.\par

Integrating the fermion loop as before \cite{Ball} and
renormalizing appropriately $W_\mu$, we get  at the scale
$\Lambda_{NP}$
\bq
\L_{NP}  = \frac{(2I+1)I(I+1) \widetilde{N}_c}{ (4\pi)^2}~
 \frac{g^2_2}{45 \Lambda^2_{NP}}\left [
  g_2 \O_W - \, \ol{\O}_{DW}
\right]  \ \ . \
\eq
Comparing  with (23, 27-30), we conclude that there is no $\O_{UW}$
contribution in this model, and
\bq
\lw = ~\frac{(2I+1)I(I+1)\widetilde{N}_c}{45\ (4\pi)^2}~
\frac{g^2_2 \mwd}{\Lambda^2_{NP}} \ \ \ \ \ , \ \ \ \
 \eq
\bq
x_{\ol{DW}} ~=~ -\ \lw \ \ \ \ \ \ \ \ . \ \ \ \ \ \ \
\eq\par

Several other nontrivial features can also be noticed. Thus,
the sign of $\lw$ is now
positive, opposite to that of $x_{\ol{DW}}$ and opposite to $\lw$ in
model A. Moreover $\O_W$ and $\ol{\O}_{DW}$ are now generated
with equal strength. As before the actual magnitude of the
couplings is controlled
by the isospin and hyper-colour factors.\par

At this stage no NP couplings to the Higgs field are generated in
model B. In order to get them the model is extended to\par
\vspace{0.5cm}
\noindent
\newpage
\underline{Model B$^\prime$ :}\par
To the preceding NP spectrum we just add a heavy scalar
isoscalar field $S^0$. The most general gauge invariant and
renormalizable lagrangian
is now given by (32), with (33) with $\L_F$ given  by
\bq
\L_F=~i \ol{F} (\rlap/\partial +i g_2 \rlap /W )F
-M_f \ol{F}F +f_f S^0\ol{F}F+ \frac{f_\phi}{2} M_s S^0\langle
\Uh \Uh^\dagger
\rangle +\frac{1}{2}(\partial_\mu S^0)(\partial^\mu S^0) -
\frac{M^2_s}{2}\, (S^0)^2   \ ,  \
\eq
where\footnote{Irrelevant terms like $(S^0)^4$ and
$(S^0)^2\langle \Uh \Uh^\dagger \rangle$ are omitted in (37), and
the vacuum expectation value of $S^0$ is assumed zero.}
the $S^0$ and $F$ masses are
assumed to be similar; \ie\ $M_s\sim M_f \equiv \Lambda_{NP}$.
So essentially only two additional parameters have been
introduced in this extension, namely the dimensionless couplings
$f_f$ and $f_\phi$. A priori, they are also of order 1.\par

By integrating the heavy fermion loop as before \cite{Ball},
we obtain the effective lagrangian, which as far as the $\O_W$
and $\ol{\O}_{DW}$ contributions are concerned, is the same as the
one given in (34). In addition to it though, the heavy fermion
loop also creates an effective $WWS$ coupling of the form
\bq
\L_{SWW}~=~ \frac{x_s}{M_s} \overrightarrow{W}_{\mu \nu}
\overrightarrow{W}^{\mu \nu}\ S^0 \ \ \ \ \ , \ \ \ \
\eq
\bq
x_s~=~-\ \frac{g^2_2\widetilde{N}_c}{(4 \pi)^2} ~
\left (\frac{2f_f I(I+1)(2I+1) M_s}{9 M_f}
\right )\ \ \ \ \ . \ \ \ \ \
\eq
The $S^0$ exchange between a pair of Higgs doublets and a W pair
then generates  $\op{\O}_{UW}$ with a coupling (compare (23))
\bq
\tilde d ~=~ -\frac{f_f f_\phi \widetilde{N}_c}{18 \pi^2}
\left (\frac{\mwd}{M_f M_s} \right) I(I+1)(2I+1)\ \ \ \ \ \ \ \ .\ \ \ \
\eq
Note that the size of $\tilde d$ depends on the unknown
factor  $f_f f_\phi$. Comparing to the $\O_W$ case,
we remark that if this factor is of the order of 1, or even
of the electroweak order $g^2_2$, then we expect that
$| \tilde d / \lw | \gsim 30$; (compare (35,40)).\par

As we see from the expressions for the various coupling
constants, the above models illustrate how NP can generate
a possibly strong selection among the induced operators. This
selection  can appear as a result  of the chosen spectrum
in NP and of the precise dynamics determining the size of the
various couplings. It may  even take the
aspect of a symmetry. Whether it really corresponds to a
basic symmetry of NP is an open question. This way, NP can
prohibit the generation of unwanted large
$\Delta \rho$ or $W-B$ mixing effects, ruled out by precise
tests. \par

We also want to make several remarks about the size of the
effects. In models A and B$^\prime $ we can naturally
accommodate larger
couplings for the  Higgs involving operators, than
for those  involving $W$ fields alone; \ie\@
$(\O_{\Phi 2,3} \gsim \O_{UW} \gsim \O_{W}) $. Of course, this
is only true provided scalar elementary NP bosons exist.
Otherwise the Higgs involving operators vanish. The
couplings of the purely gauge dependent operators are rather weak.
This weakness most probably arises
from the fact that they solely depend on gauge couplings,
for which there is no
freedom. On the contrary, the Higgs involving operators
naturally share the well known arbitrariness of the scalar sector
of SM. And it is this lack of knowledge for the scalar sector,
which makes
operators like $\O_{UW}$ very interesting. We also repeat that
in model A, the $SU(2)_c$ violating coupling
$g_{\psi 2}$ is experimentally constrained to be very small. \par

Turning now to the purely gauge operators above, we note that the
$\O_W/ \ol{\O}_{DW}$ ratio is very model
dependent in sign and magnitude. As mentioned before, the
physical contents of these operators is very different.
$\O_W$ includes no 2-pt function, and its 3-pt function
defines the genuine quadruple gauge boson
coupling $\lw$ \cite{Kuroda}.
$\ol{\O}_{DW}$, on the other hand, has 2-pt and 3-pt gauge vertices
which are just
d'Alembertien derivatives of the standard kinetic terms. For
this reason the $\ol{\O}_{DW}$ coupling does not seem to bring
much NP feature. It is
thus conceivable that $\O_W$ and $\ol{\O}_{DW}$ may get very
different NP effects.\par

 It is well known that present precision
measurements strongly constrain $\ol{\O}_{DW}$ through its 2-pt
function part. Do these already exclude our models A and B?
We argue below that this is not necessarily the case,
since non-perturbative effects due to possible existence of
NP resonances could avoid the conclusion that
$\O_W$ and $\ol{\O}_{DW}$ have comparable couplings.
As an example we consider the possibility of vector
bosons $V$, which are non-relativistic bound states of
either the $\Psi$ particles in  model A, or the
$F$ particles in model B. The  gauge boson 2-pt function
receives then a contribution
from W-V mixing, which involves the V wave function or some
of its derivatives
at the origin. This effect should have the standard SM
structure, and therefore it should be associated to
$\ol{\O}_{DW}$. On another hand the quadruple 3-boson coupling
described by $\O_W$ receives a
contribution involving an integral over the full V wave function.
Depending  on the relation between the average
radius $\bar r$ of the $V$ wave function and $1/\Lambda_{NP}$,
different situations on the $\O_W/ \ol{\O}_{DW}$ ratio may
 arise. Thus, in case $\bar r \gg 1/\Lambda_{NP}$,
we would expect that both, the $\ol{\O}_{DW}$ and the $\O_W$
couplings depend on the short distance behaviour of the V wave
function. In such a case it may be natural to expect these couplings
to be similar.  Another more interesting
scenario may arise  in case $\bar r \lsim 1/\Lambda_{NP}$
since then $\O_W$ depends on the full wave function, which may in
turn imply that  $\O_W$ becomes
 very different from $\ol{\O}_{DW}$. Thus, the present strong
constraints on $\ol{\O}_{DW}$, should not be taken as excluding
the possibility of considerable $\O_W$ contributions
in future experiments.\par

In this paper we have presented dynamical pictures where NP generates
at a scale  $\Lambda_{NP}$
a hierarchy of mainly $SU(2)_c$ invariant operators. The basic
characteristic of our models is the assumption that the new degrees
of freedom determining NP have vanishing hypercharge.
The natural leading terms in the induced effective lagrangian
involve operators containing Higgs fields; \ie\@
($(\O_{\Phi 2,3}$ and $\O_{UW}$). Whether $\O_W$
is enhanced with respect to the strongly constrained
$\ol{\O}_{DW}$, is a more model dependent question.
In model A, an $SU(2)_c$ violating operator is also generated at
$\Lambda_{NP}$, determined by an independent coupling
$g_{\psi 2}$ which is constrained by
present data to be very small. On the contrary, in Model B only
$SU(2)_c$ conserving operators appear. It is also worth
mentioning that if the hypercharge of the $\Psi$ or $F$ NP particles
were non vanishing, then the operators $\O_{DB}$, $\O_{BW}$ and $\O_{UB}$
would also be generated in model A, while in model B only
$\O_{DB}$. The operators $\O_{W\Phi}$ and $\O_{B\Phi}$ never
appear in such models at the level of $dim=6$ operators. In type
A models, $\O_{W\Phi}$ or $\O_{B\Phi}$, multiplied by
$\Phi^\dagger \Phi$,   first arise at the
$dim=8$ level . \par

 The overall picture
is consistent with the view that NP has something to
do with the scalar sector. The non
observation of some NP effect in the light fermionic sector
at LEP1, does not prevent potentially large NP effects to appear
in the bosonic sector. These features should be tested at future
colliders through gauge boson and Higgs production. They would
then provide valuable information on the New Physics
properties and its possible origin.\par
\vspace {0.5cm}
\noindent{\large \bf Acknowledgements}\par
 One of us (F.M.R.) wishes to thank the Department of Theoretical
 Physics of the
 University
of Thessaloniki for the warm hospitality
as well as the kind help that he
 received
 during his
stay and for the creative atmosphere that was generated during
the preparation of this paper.

\end{document}